\documentclass[aps,twocolumn,prl,showpacs,amsmath,amssymb,10pt]{revtex4-1}
\usepackage{graphicx}
\usepackage{scrextend}
\usepackage{manfnt,pifont}
\usepackage{hyperref}
\usepackage{url}
\usepackage[outline]{contour}
\usepackage{textgreek}

\usepackage{CJK}
\usepackage{amsmath,amssymb,lmodern}
\usepackage{verbatim}
\usepackage{hyperref}
\usepackage{physics}
\usepackage{amsfonts}
\usepackage{amssymb}
\usepackage{bbold}

\begin{document}

\begin{CJK*}{GB}{}
\CJKfamily{gbsn}

\title{Random Bond perturbations of the $O(2)$ vector model}

\author{Maria Nocchi}
\affiliation{Mathematical Institute, University of Oxford, Andrew Wiles Building, Radcliffe Observatory Quarter, Woodstock Road, Oxford, OX2 6GG, U.K.}

\begin{abstract}
\noindent We investigate the impact of quenched disorder in the critical $O(2)$ vector model. We first review, in the modern language of Conformal Perturbation Theory, the random temperature perturbation in $4-\epsilon$. Then, we present a direct computation in $3d$. The pure CFT is now strongly interacting, and its CFT data are determined in recent numerical bootstrap studies. We then explore the anisotropic disorder associated with the lowest-dimension charge-$2$ scalar, which is relevant in $3d$. This perturbation breaks locally $O(2)$. However, the symmetry is restored in the IR in a weakly coupled new fixed point.
\end{abstract}

\maketitle
\end{CJK*}

\section{Introduction}
\noindent
\label{sec:intro}

The Renormalization Group (RG) stands as a powerful framework in Theoretical Physics, providing a systematic way to understand the behavior of theories under scale transformations. Particularly notable are systems invariant under such transformations, constituting fixed points of the RG flow. Remarkably, most scale-invariant physical theories exhibit conformal invariance  \cite{Nakayama:2013is}. 

The RG procedure generates a scale-dependent effective action characterized by a set of local operators classified as irrelevant, relevant, or marginal. Regardless of the specific combination of fields (and their derivatives) allowed by the symmetries, theories with possibly very different microscopic descriptions may converge to the same infrared (IR) fixed point, corresponding to a Conformal Field Theory (CFT). This makes manifest the underlying universality of physical phenomena.

When discussing statistical or condensed matter systems, Hamiltonians are typically defined on regular lattices. While this description concerns idealized homogeneous systems, realistic systems often present local random impurities or non-constant background fields. As a result, the microscopic interactions occurring inside the substance are modified. Therefore, extending the standard RG apparatus to disordered systems is of great interest. 

The couplings become random functions of position with a fixed probability distribution. Here, we focus on the case of quenched disorder, where the random variables describing the disorder fluctuate over timescales considerably larger than observable times \cite{brout1959statistical, cardy1996scaling}. In this sense, the disorder is frozen and the impurities act as non-homogeneous background fields. We are not interested in analyzing such systems for specific realizations of random disorder, where translation invariance is broken. Instead, we study the statistical properties by averaging the free energy $\log Z$, with $Z$ being the partition function, over the disorder. 

The big goal is to understand whether the fixed point corresponding to the pure system is stable against disorder. The potential emergence of a new critical behavior hinges on whether the disorder perturbation is relevant or marginal. If this is the case, even a small concentration of impurities can be important under the RG flow \cite{Narovlansky:2018muj} and the critical exponents will differ from the ones in the absence of disorder. The Harris criterion \cite{Harris:1974zza}, relying on the knowledge of the critical behavior of the pure system only, will be a predictive tool in this context. 

As a concrete standard example, one can consider the Ising Model and insert a quenched disorder associated either with the energy operator, known as the \textit{random bond} case \cite{Komargodski:2016auf}, or the spin operator, corresponding to the so-called \textit{random field} model \cite{Kaviraj_2020, Kaviraj_2021}. 
In this work, we further pursue this research direction and present the effect of introducing a quenched disorder in the $O(2)$ vector model. In particular, we analyze two kinds of random perturbations: one preserves the global symmetry, akin to the Random Bond Ising case, while the other one is charged under $O(2)$, breaking the symmetry locally. This dual perspective enriches our understanding of how disorder affects the phase transitions and critical behavior of systems with continuous symmetry. 

\section{Replica Trick and
Conformal Perturbation Theory}

In the field-theoretic language, the disorder is introduced by perturbing the pure action with a random coupling, denoted as $h(x)$, associated with the local operator $\mathcal{O}(x)$ which tunes the system across the transition:
\begin{equation} \label{93}
S_{\rm{CFT}} \to S_{\rm{CFT}} + \int d^d x \ h(x) \mathcal{O}(x) \ .
\end{equation}
Here $h(x)$ is a fixed random function dependent only on space. An efficient approach to compute the averaged disordered free energy is through the Replica Trick (\cite{kac2013certain}, \cite{edwards1971statistical}, \cite{ma1972critical}), which originates from the identity: 
\begin{equation}
F \sim \log Z = \lim_{n \to 0} \frac{Z^n -1}{n}= \lim_{n \to 0} \frac{\partial Z^n }{\partial n} \ .
\end{equation}
This way, the free energy is rewritten in terms of the power of the partition function, which itself can be read as a partition function of a different theory ($n$ copies of the original theory).
The disorder-induced interaction takes the form
\begin{equation} \label{95}
g \int d^d x \ \sum_{A,B=1}^n \mathcal{O}_A(x) \mathcal{O}_B(x) \ ,
\end{equation}
where $n$ is  thought as a positive integer here. Then, we will work with analytic expressions in $n$ and send it to $0$ to get the physical theory. 

In \eqref{95}, the new coupling $g$ no longer depends on spatial coordinates. This framework is suitable for applying Conformal Perturbation Theory (CPT) to study the evolution of the coupling induced by the disorder. CPT tells us how to expand around a non-trivial fixed point ~\cite{cardy1996scaling} by knowing only the CFT data of the unperturbed CFT, which we refer to as the pure system. We start from the initial fixed point Hamiltonian $H^*$ and write the perturbed partition function as
\begin{equation} \label{49}
Z = \Tr e^{-H^* + \sum_i g_i \int d^d x \ \mathcal{O}_i(x)} \ .
\end{equation}	
Then, we expand it in powers of the couplings.
Following \cite{Komargodski:2016auf}, we extract the beta-functions from the overlaps $\braket{\mathcal{O}_m}{0}_{g,V}$, where
\begin{equation} \label{13}
\begin{split}
    & \bra{\mathcal{O}_m} \equiv \bra{0} \mathcal{O}_m(\infty) \ , \\
    & \mathcal{O}_m(\infty) \equiv \lim_{x \to \infty} x^{2 \Delta_{\mathcal{O}_m}} \mathcal{O}_m(x) \ , \\
    & \ket{0}_{g,V} \equiv e^{\sum_i g_i\int_V d^d x \ \mathcal{O}_i(x)} \ket{0} \ .
\end{split}
\end{equation}
Here $g$ stands for the set of the couplings $\{g_i\}$ and $V$ is a volume around the origin in which we deform the CFT. We do this to handle IR divergences. Putting it all together:
\begin{equation} \label{52}
\begin{split}
& \braket{\mathcal{O}_m}{0}_{g,V} \\
& \sim V g_m + \frac{1}{2} V \sum_{i,j} g_i g_j \int d^d x \ \langle \mathcal{O}_i(0) \mathcal{O}_j(x) \mathcal{O}_m(\infty)\rangle \\
& + \frac{1}{6} V
\sum_{i,j,k} g_i g_j g_k \int d^d x \ d^d x_1 \ \langle \mathcal{O}_i(0) \mathcal{O}_j(x) \mathcal{O}_k(x_1) \mathcal{O}_m(\infty)\rangle \\
& + \dots \ .
\end{split}
\end{equation}
The correlation functions are evaluated with respect to the fixed point Hamiltonian and computed by using the Operator Product Expansion (OPE):
\begin{equation} \label{12}
\mathcal{O}_i(x) \times \mathcal{O}_j(0) \sim \frac{\delta_{ij}}{|x|^{2 \Delta_{\mathcal{O}}}} + \sum_k \frac{C_{ijk}}{|x|^{\Delta_{\mathcal{O}_k}}} \mathcal{O}_k(0) + \dots \ ,
\end{equation}
where $\Delta_{\mathcal{O}_i}=\Delta_{\mathcal{O}_j}=\Delta_{\mathcal{O}}$ and the dots represent irrelevant scalars or operators with spin. 

We are interested in the case of marginal or slightly relevant deformations with $\Delta_{\mathcal{O}_i} = d -\delta_i$, where $\delta_i=0$ or $\delta_i \ll 1$ respectively. The integrated three- and four-point functions need to be regularized. To do so, we demand the overlap, which is a physical observable, to be UV independent of the regulator:
\begin{equation}
\mu \frac{d}{d\mu} \braket{\mathcal{O}}{0}_{g,V} = 0 \ .
\end{equation}
The resulting beta-function coefficients are proportional to the coefficients of the logarithmic terms from the integrated correlators. At one loop, this implies that
\begin{equation} \label{29}
\beta(g) = - \delta g -	\frac{1}{2} C_{\mathcal{O}\mathcal{O}\mathcal{O}} S_{d-1} g^2 + \dots \ ,
\end{equation}
where $S_{d-1} = 2 \pi^{d/2} \ \Gamma^{-1}(d/2)$. 

Notice that, when dealing with an isolated CFT, $\delta$ is actually a small fixed number. The usual strategy is to pretend that a family of CFTs with $\delta \to 0$ does exist, although the CFT data is known only at a fixed $\delta \equiv \tilde{\delta}$. In our study, this is the case of three-dimensional systems, where the CFT data has been derived in numerical bootstrap studies. Therefore, the results are trusted only at the leading non-trivial order, as the OPE coefficients are defined with an error $O(\tilde{\delta})$. The first order of the beta-function in which they enter is thus universal, but the next one is no longer well-defined, requiring a case-by-case understanding of how to proceed.

To conclude this review section, let us mention how to compute anomalous dimensions in CPT. Inserting a local perturbation in a theory results in the other operators in the spectrum acquiring anomalous dimensions. 
For each operator $\mathcal{A}$, its anomalous dimensions in terms of a power expansion in the disorder-induced coupling $g$ is
\begin{equation}
    \Delta_{\mathcal{A}}(g)^{\rm{IR}} =  \Delta_{\mathcal{A}}^{\rm{UV}} + \gamma_{\mathcal{A},1} g + \dots \ , 
\end{equation}
with the one-loop correction being
\begin{equation}
    \gamma_{\mathcal{A},1} = - S_{d-1} C_{\mathcal{A} \mathcal{A} \mathcal{O}} \ .
\end{equation}
The key takeaway is that by using the Replica Trick and the formalism of Conformal Perturbation Theory, we can extract the properties of random fixed points describing second-order phase transitions in disordered materials. For instance, anomalous dimensions tell us about the (new) critical exponents. 
%
\section{Isotropic Random Bond $O(2)$ Model}
\label{sec:IRBOM}
The $O(N)$ theory is the simplest generalization of the Ising-model with $N$-component unit-length spins $\{\vec{s}_i\}$. At criticality, the system is described by a CFT and a large collection of conformal data is available by now \cite{Henriksson:2022rnm}. 
Here we focus on the $N=2$ case. The $O(2)$ universality class describes several phenomena such as the superfluid transition in $^4$He (the so-called $\lambda$-point experiment \cite{PhysRevB.68.174518}), the transitions in magnets with easy-plane anisotropy, the formation of Bose-Einstein condensates, etc. In the field-theoretic description, one classifies the low-lying operators in terms of the irreducible representations of the global $O(2)$: the order parameter is a scalar field in the fundamental; the energy field ($s=\sum_i \phi_i \phi_i$) is a singlet, and here we will also consider the symmetric traceless tensor ($t_{ij}=\phi_i \phi_j-{\rm trace}$).

The leading scaling dimensions and the OPE coefficients for the $3d$ $O(2)$ model are known from numerical bootstrap studies \cite{Chester:2019ifh}. We report the CFT data we need in the present work \footnote{The estimates for the dimensions are exact: the numbers in the brackets are not a confidence interval, but rather a true limit, while ($^*$) stands for non-rigorous estimates (see the original article for the details).}:
\begin{equation} \label{94}
\begin{split}
&\Delta_s = 1.51136(22) \ , \\
& \Delta_\phi = 0.519088(22) \ , \\
&\Delta_t = 1.23629(11) \ .
\end{split}
\end{equation}
We first show the consequences of introducing a (ferromagnetic) random bond perturbation, which we refer to as the isotropic case. This may correspond to non-magnetic vacancies, for instance. We reproduce the well-known result that the critical behavior is modified at first order in $\epsilon$-expansion around $4d$ (see \cite{PhysRevB.15.258} and references therein). It is always the case that two non-trivial fixed points are found but only one is stable for any $N$: for $N>4$ the Wilson-Fisher point is stable, while the random fixed point is stable for $N<4$. We then present the (direct) $3d$ computation. Lastly, we extend the analysis to a quenched disorder associated with $t_{ij}$. 

We are not going to discuss the effect of quenched random fields in $O(N)$ models, but we refer to \cite{PhysRevLett.35.1399,PhysRevLett.37.944} for details on this and useful additional references. This disorder generates the random bond disorder under RG flow and one needs to consider both. 

Let us now go into detail about the Random Bond $O(2)$ model, described by the insertion of the singlet $s(x)$ with a random coupling $h(x)$, taken to be Gaussian with zero mean and variance $c^2$. It is also conventional to assume the random variables to be uncorrelated for different bonds (or sites). As anticipated, we use the Replica Trick to build a set of $n$ identical decoupled CFTs, each of them coupled to the same disorder. The replicated partition function is 
\begin{equation} \label{62} 
\begin{split}
    W_n = & \int Dh \prod_{A=1}^{n} D\vec{\phi}_A \ {\rm exp} \bigg (-\sum_A S_A + \int d^d x \ h(x) s_A(x) \\
    & -\int d^d x \ \frac{h(x)h(x)}{2c^2} \bigg ) \ ,
\end{split}
\end{equation}
from which the averaged disorder-free energy will be
\begin{equation}
    F_D = \frac{d}{dn} \eval{W_n}_{n=0} \ .
\end{equation}
The Gaussian Path Integral over $h$ induces the non-linear interaction
\begin{equation} \label{58}
\sum_{A,B} s_A(x) s_B(x) \ ,
\end{equation}
with coupling $c^2/2$. Since $[c^2]=d-2\Delta_s$, the interaction is irrelevant as long as $\Delta_s > d/2$ or marginal in case equality holds. Hence, in $d=3$ it is irrelevant already at the UV fixed point. We expect the RG flow to be crushed to $0$ in the IR and the pure fixed point to be stable under random temperature perturbations. However, it is marginally irrelevant: we want to check if quantum corrections compensate for this. 

We first present the $4d$ analysis, where the deformation is marginal and the computation is exact, controlled by $\epsilon=4-d$. Then we move to the $3d$ case.

\subsection{Four dimensions}
\begin{figure} 
    \centering
    \includegraphics[width=0.3\textwidth, height=0.3\textheight, keepaspectratio]{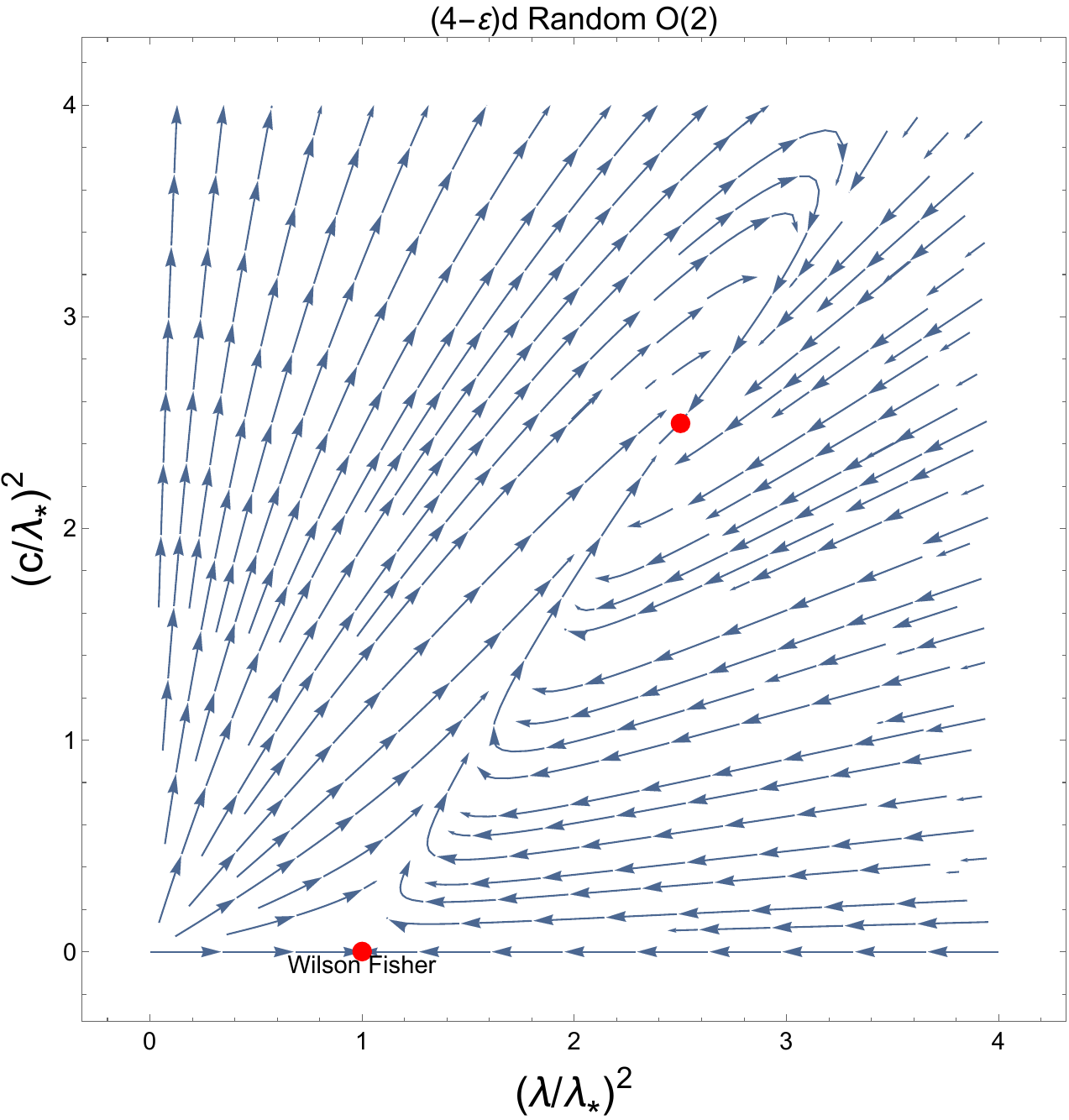}
    \caption{Phase diagram of Random Bond $O(2)$ Model in $d=4-\varepsilon$.}
    \label{figure:r5}
\end{figure}

As a first step, we write the OPE in the pure theory, that has the $O(2)$ global symmetry. The two-point function of $\phi_i(x)$ is diagonal in the $O(2)$ indices because of unitarity. Hence we define the theory starting from 
\begin{equation}
\langle \phi_i(x_1) \phi_j(x_2) \rangle = \frac{\delta_{ij}}{|x_{12}|^{2\Delta_\phi}} \ .
\end{equation}
The three-point function $\langle \phi \phi \phi \rangle$ is trivially zero because the theory is even under parity. The three-point function involving two $\phi$-operators and a scalar field is instead
\begin{equation} \label{61}
	\langle \phi_i(x_1) \phi_j(x_2) \mathcal{O}(x_3) \rangle = \frac{\delta_{ij} C_{\phi\phi\mathcal{O}}}{|x_{12}|^{2\Delta_\phi-\Delta_{\mathcal{O}}} |x_{13}|^{\Delta_{\mathcal{O}}} |x_{23}|^{\Delta_{\mathcal{O}}}} \ .
\end{equation}
The OPE of the fundamental field can be therefore written as  
\begin{equation}
\phi_i(x) \times \phi_j(0) \sim \frac{\delta_{ij}}{|x|^{2\Delta_\phi}} + \sum_{\mathcal{O}} C_{\phi\phi\mathcal{O}} (x,\partial) \mathcal{O}_{ij} (0) \ ,
\end{equation}
where the operators in the sum can be $O(2)$ singlets, antisymmetric $O(2)$ tensors or symmetric traceless $O(2)$ tensors. Similarly, we can deduce the OPE of $s$-operators that will involve only singlet operators, for instance scalars $\{s,s',s'',\dots\}$ of increasing order of dimension. In four dimensions, $\Delta_s = 2$ and $s'=s^2$ (with coupling $\lambda^2$) is marginal. We need to evaluate possible mixing under RG flow. This is done by applying the formalism of CPT in the case of more deformations, with OPE coefficients
\begin{equation} 
\begin{split}
& s(x) \times s(0) \sim \frac{1}{|x|^4} + \frac{2}{|x|^2} s(0) + s^2(0) + \dots \ , \\
& s(x) \times s^2(0) \sim \frac{4}{|x|^4} s(0) + \dots \ , \\
& s^2(x) \times s^2(0) \sim \frac{4}{|x|^8} + \frac{20}{|x|^4} s^2(0) + \dots \ .
\end{split}
\end{equation}
Another important observation is that in $4d$ the disorder-induced interaction \eqref{58} splits into two terms (sums over equal or different replicas) that transform differently under the permutation symmetry of the replicas. In general, we need a UV cutoff to regularize the product of the same operator at coincident points. We will fix this by replacing this term with the (relevant) operators allowed in its OPE. As stated above, these operators have to be considered once we deform the pure CFT, as the RG procedure would generate them \footnote{Notice that if the pure CFT is a free theory (as in the four-dimensional cases here), $\mathcal{O}^2$ appears in a non-singular way in the OPE.}. Let us study \begin{equation} \small
\begin{split}
    W_n = & \int \prod_{A=1}^{n} D\vec{\phi}_A \ {\rm exp} \bigg (-\sum_A S_A + \int d^4 x \ (-\lambda^2 + \tfrac{c^2}{2}) \sum_A s^2_A  \\
    & + \tfrac{c^2}{2}\sum_{A\ne B}s_A s_B \bigg )  \ .
\end{split}
\end{equation}
We will denote the two interactions by
\begin{equation}
\begin{split}
& \mathcal{O}_1 \equiv \frac{1}{\sqrt{4n}} \sum_A s_A^2 \ , \\ 
& \mathcal{O}_2 \equiv \frac{1}{\sqrt{2n(n-1)}} \sum_{A \ne B} s_A s_B \ , 
\end{split}
\end{equation}
and their couplings by 
\begin{equation}
\begin{split}
& g_1 \equiv \sqrt{4n} ( -\lambda^2 + \tfrac{c^2 }{2} ) \ , \\
& g_2 \equiv \sqrt{2n(n-1)} \ \tfrac{c^2}{2} \ .
\end{split}
\end{equation}
The procedure is to compute the three-point functions in the replicated theory, plug them into the one-loop beta functions, and take $n \to 0$ at the end. This leads to
\begin{equation}
\begin{split}
& \frac{d\lambda^2}{d\log \mu} = 2 S_3 \lambda^2 (5 \lambda^2 - 3 c^2) + \dots  \ , \\
& \frac{dc^2}{d\log \mu} = 4 S_3 c^2 (2\lambda^2 - c^2) + \dots \ .
\end{split}
\end{equation}
It is clear that $\lambda^2$ is irrelevant in the absence of disorder. Moreover, if $\lambda^2=0$, the theory at the beginning flows to strong coupling. This is the same situation of the Ising model \cite{Komargodski:2016auf}. However, the one-loop approximation has no line of fixed points. We therefore expect that in $d=4-\varepsilon$ there could be solutions of order $\varepsilon$. Indeed, the beta-functions acquire a tree-level term $-\varepsilon g_i$, and for the standard case ($c^2=0$), one can read off the usual free fixed point and the Wilson-Fisher one:
\begin{equation}
\bigg (\lambda^2_*,c^2_* \bigg ) = \bigg (\frac{\varepsilon}{10 S_3},0 \bigg ) \ .
\end{equation}    
But, contrary to the Ising case, there is another fixed point:
\begin{equation}
\bigg (\lambda^2_*,c^2_* \bigg ) = \bigg (\frac{\varepsilon}{4 S_3},\frac{\varepsilon}{4 S_3} \bigg ) \ .
\end{equation}    
In the Random Bond isotropic $O(2)$ model close to $4d$ there are two perturbative fixed points: the Wilson-Fisher point, which is unstable, and the disorder-induced new fixed point which is stable. The phase diagram can be seen in figure \ref{figure:r5}.

We end this section by computing the anomalous dimension of the energy operator. If we define the replicated energy operator,
\begin{equation}
    S(x) \equiv \frac{1}{\sqrt{n}} \sum_{A=1}^n s_A(x) \ ,
\end{equation}
we find 
\begin{equation}
    C_{SS\mathcal{O}}=\frac{2(n-1)}{\sqrt{2n(n-1)}} \ ,
\end{equation}
which gives
\begin{equation}
    \eval{\Delta_S^{\rm{IR}} (c^2)}_{n=0} = 2 + S_3 c^2 + \dots \ . 
\end{equation}
At the disordered fixed point, the one-loop anomalous dimension of the energy operator is 
\begin{equation}
    \gamma_{s,1} = \frac{\epsilon}{4} \ . 
\end{equation}

\subsection{Three dimensions}
The case of the $3d$ $O(2)$ model is simpler because it is believed that $s'$ is irrelevant in the unperturbed pure CFT ($\Delta_{s'} = 3.789(4)$ \cite{hasenbusch2019monte}). Notice that the operator $t_{ij}$ is relevant in three dimensions \eqref{94}, but it does not appear in the composition of two singlets. The $3d$ replicated partition function is 
\begin{equation}
W_n = \int \prod_{A=1}^{n} D\vec{\phi}_A e^{-\sum_A S_A + \frac{c^2}{2} \int d^d x \ \sum_{A \ne B} s_A(x) s_B(x)} \ ,
\end{equation}
where $s_A$ has dimension close to $d/2$, with $\delta \equiv d-2\Delta_s \simeq -0.02$. The one-loop beta-function for the strength of disorder is
\begin{equation} \label{90}
\frac{dc^2}{d \log \mu} \simeq 0.02 c^2 + S_2 1.65 c^4 + \dots \ .
\end{equation}
If $c^2 \ll 1$ the disorder is marginally irrelevant in $d=3$: in this case, the Random Bond $O(2)$ Model is similar to the $2d$ Ising model with a random bond magnetic field \cite{pleimling2004logarithmic}.

To sum up, the random temperature perturbation for $O(2)$ is relevant in $\varepsilon$-expansion, but becomes irrelevant in $d=3$. It would be interesting to explore this observation further, for example using resummation techniques. We leave this as a future direction of research.

\section{Anisotropic Random Bond $O(2)$ Model}
\label{sec:ARBOM}

In the (pure) $O(N)$ vector models with $N>1$, exploring perturbations that break the global symmetry becomes crucial. The stability properties of $O(N)$ vector models, concerning fixed points such as cubic, biconical, or decoupled $O(n_1) + O(n_2)$, prompt the examination of anisotropic perturbations to understand their impact on critical behavior. The lowest sector of $O(N)$ spin-$l$ operators often yields the most significant effects  \cite{hasenbusch2011anisotropic}. 
Focusing on the specific case of the $3d$ $O(2)$ model, it is found to be unstable against the anisotropy of $l = 2$, corresponding to the insertion of $t_{ij}$. From a Hamiltonian perspective, if we consider the insertion of $t_{11}$ only, this perturbation is equivalent to inserting opposite mass terms for the two components of the order parameter. 

Our study delves into the consequences of such perturbations when associated with a random coupling. Specifically, we consider the insertion of $t_{ij}$, which is relevant in $d=3$ \cite{Chester:2019ifh}, with a random coupling $h_{ij}(x)$. 

Let us briefly review the spectrum of the pure theory. In the usual bootstrap approach, assumptions include the presence of one relevant singlet scalar $s$, one relevant vector scalar $\phi_i$, and one relevant traceless symmetric scalar $t_{ij}$. Then there are gaps to the second-lowest dimension operators in each sector. It is believed that $s'$, $\phi'$, and $t'$ are instead irrelevant. Moreover, for the second charge-$2$ operator, $\Delta_{t'} = 3.624(10)$ \cite{Calabrese:2002bm}. 

For the three-point functions, as seen before, the ones involving only $\phi$-operators are zero, from the symmetry $\mathbb{Z}_2 \subset O(2)$. Similarly, from $SO(2) \subset O(2)$, the three-point function of $t_{ij}$ is zero as well. 
This results in the OPE 
\begin{equation} \label{64}
t \times t \sim \mathbb{1} + s + \dots \ .
\end{equation}
The disordered free energy is now expressed as
\begin{equation}
F_D = \int Dh_{ij}(x) F[h_{ij}(x)] e^{-\frac{1}{2}\int d^d x \ h_{ij}(x) M_{ijkl} h_{kl}(x)} \ ,
\end{equation}
where $M_{ijkl}$ is interpreted as the analogue of $1/c^2$ in the case of the random temperature perturbation \eqref{62}. Indeed, this four-index tensor contains information about the Gaussian distribution widths of the two independent components of $t_{ij}$. As before, the Gaussian Path Integral over $h_{ij}$ generates
\begin{equation} \label{91}
\sum_{A,B} t^A_{ij}(x)t^B_{kl}(x) \ .
\end{equation}
Terms with $A = B$ are discarded, as the operator $\sum_A t_{ij}^A t_{ij}^A$ is irrelevant in $3d$. We could wonder whether to consider the operator $\sum_A s_A$, given that $s$ is relevant in $3d$. However, we assume we tuned the temperature to $0$ and focus on
\begin{equation} \label{60}
\mathcal{O}_{ijkl} \equiv \sum_{A \ne B} t^A_{ij}(x)t^B_{kl}(x) \ .
\end{equation}
The next step is to get the OPE coefficients in the replicated theory. The two-point function of $t_{ij}$ in the unperturbed theory is
\begin{equation}
\langle t_{ij}(0) t_{kl}(x) \rangle = \frac{T_{ijkl}}{|x|^{2\Delta_t}} \ ,
\end{equation}
where $T_{ijkl} \equiv \delta_{ik} \delta_{jl}+\delta_{il} \delta_{jk}-\delta_{kl} \delta_{ij}$. Then:
\begin{equation} \label{70} \small
\begin{split}
& \langle \mathcal{O}_{ijkl} (0) \mathcal{O}_{mnpq}(x) \rangle = \frac{n(n-1)}{|x|^{4\Delta_t}} (T_{ijmn} T_{klpq} + T_{ijpq}T_{klmn} ) \ , \\
& \langle \mathcal{O}_{ijkl}(0) \mathcal{O}_{mnpq}(x) \mathcal{O}_{rsuv}(y) \rangle  = \frac{n(n-1)(n-2)}{|x|^{2\Delta_t}|y|^{2\Delta_t}|x-y|^{2\Delta_t}} \\
& \times (T_{ijmn} T_{klrs} T_{pquv} + \rm{perms} ) \ .
\end{split}
\end{equation}
Moreover, from the symmetry and tracelessness of $t_{ij}$, the components of $\mathcal{O}_{ijkl}$ are reduced to three independent entities: ${O_{1111} \equiv O_1, O_{1212} \equiv O_2, O_{1222} \equiv O_3}$.

\subsection{Three dimensions}
In $3d$, the disorder associated with $t_{ij}$ is relevant and induces a rank-four perturbation in the replicated theory. The fixed-point replicated action is deformed according to  
\begin{equation} \label{69}
S \to S + \sum_{ijkl} g_{ijkl} \int d^d x \ \mathcal{O}_{ijkl}(x) \ ,
\end{equation}
where $g_{ijkl} = M^{-1}_{ijkl}/2$. 
In our case, the symmetries of the theory lead to a significant simplification and the sum in \eqref{69} has just three independent terms, with couplings:
\begin{equation}
    \begin{split}
        & g_1 \equiv g_{1111}+g_{2222}-2g_{1122} \ , \\
& g_2 \equiv 4 g_{1212} \ , \\
& g_3 \equiv 4(g_{1222}-g_{2111}) \ .
    \end{split}
\end{equation}
Given that $\Delta_t = \tfrac{d-\delta}{2}$, with $\delta \simeq 0.53$, the beta-functions are
\begin{equation}
\begin{split}
& \beta_{g_1} = - 0.53 g_1 + 4 S_2 g_1^2 + S_2 g_3^2 + \dots \ , \\ 
& \beta_{g_2} = - 0.53 g_2 + 4 S_2 g_2^2 + S_2 g_3^2 + \dots \ , \\
& \beta_{g_3} = - 0.53 g_3 + 4 S_2 g_1 g_3 + 4 S_2 g_2 g_3 + \dots \ .
\end{split}
\end{equation}
Notice that $g_1$ and $g_2$ correspond to interactions that break $O(2)$ to $\mathbb{Z}_2$, while $g_3$ breaks also $\mathbb{Z_2}$. From the above beta-functions, we consider the following fixed points:
\begin{equation}
    (g_1^*,g_2^*,g_3^*)=\left \{\left (\tfrac{0.53}{4 S_2},0,0 \right ),\left(0,\tfrac{0.53}{4 S_2},0 \right ),\left(\tfrac{0.53}{4 S_2},\tfrac{0.53}{4 S_2},0\right) \right \} \ .
\end{equation}
Focusing on the last one with $g_1^*=g_2^*$, this corresponds to a new (random) fixed point which has $O(2)$ symmetry! This is clear if we consider that a coupling invariant under the global symmetry would be of the form $g \delta_{ik} \delta_{jl}$ (eventually symmetrized) in \eqref{69}. 

Let us also notice that the components of $g_{ijkl}$ being the variances of the components of the anisotropic disorder $h_{ij}(x)$ implies that this fixed point exists when there is a specific relation among the moments of the distribution describing this disorder. 

\subsection{Four dimensions}
To conclude, we report the analysis of our system around four dimensions. Now, the pure theory features two marginal operators: the second charge-$0$ operator
\begin{equation}
s' = \frac{1}{4} \left (\sum_i \phi_i \phi_i \right )^2 \ , 
\end{equation}
and the rank-four tensorial operator
\begin{equation}
\begin{split}
t_{ijkl} = & \phi_i \phi_j \phi_k \phi_l - \tfrac{s}{3} (\delta_{ij}\phi_k \phi_l + \delta_{ik}\phi_j \phi_l + \delta_{il}\phi_j \phi_k \\
& + \delta_{jk}\phi_i \phi_l + \delta_{jl} \phi_i \phi_k + \delta_{kl}\phi_i \phi_j)+ \tfrac{1}{6} s' (\delta_{ij}\delta_{kl} \\
& +\delta_{ik}\delta_{jl}+\delta_{il}\delta_{jk}) \ .
\end{split}  
\end{equation}
They appear in the OPE of $t$-operators:
\begin{equation} 
t_{ij} \times t_{kl} \sim \mathbb{1} + s + s' + t_{ijkl} + \dots \ ,
\end{equation}
and this means that in the replicated theory 
it is necessary to check their possible mixing with the interactions induced by disorder in the RG sense. Again, we consider the temperature tuned to $0$ and neglect the contribution of $\sum_A s_A$.

By applying CPT with the deformations 
\begin{equation} \label{2} \small
\begin{split}
& \mathcal{O}_1 \equiv \frac{1}{\sqrt{2n(n-1)}}\sum_{A \ne B}  t_{11}^A t_{11}^B \ , \\
& \mathcal{O}_2 \equiv \frac{1}{\sqrt{2n(n-1)}} \sum_{A \ne B} t_{12}^A t_{12}^B \ , \\
& \mathcal{O}_3 \equiv \frac{1}{\sqrt{n(n-1)}} \sum_{A \ne B} t_{12}^A t_{22}^B \ , \\
& \mathcal{O}_4 \equiv \frac{1}{\sqrt{4n}} \sum_A s'_A \ , \\
& \mathcal{O}_5 \equiv \frac{1}{\sqrt{3n}} \sum_A t_{1111}^A  \ , \\
& \mathcal{O}_6 \equiv  \frac{1}{\sqrt{3n}} \sum_A t_{1222}^A  \ ,
\end{split}
\end{equation} 
we find the one-loop beta-functions:
\begin{equation} \small \label{92}
\begin{split}
\beta_{g_1} =& -\varepsilon g_1 + S_3 (-4g_1^2+4g_1g_2+6g_1\lambda_5-2 g_3^2 + 3 g_3 \lambda_6 \\
& + 4 g_1 \lambda_4) \ , \\
\beta_{g_2} =& -\varepsilon g_2+S_3 (-4g_2^2+ 4g_1g_2 -6g_2\lambda_5-2g_3^2+3g_3\lambda_6 \\
& +4g_2\lambda_4 ) \ , \\
\beta_{g_3} =& -\varepsilon g_3 +S_3 (-4g_1g_3-4g_2g_3+4g_3\lambda_4+6g_1\lambda_6+6g_2\lambda_6 ) \ , \\
\beta_{\lambda_4} =& -\varepsilon \lambda_4 + S_3 ( 4g_1^2 + 4g_2^2 + 2g_3^2+5g_1g_2-6g_1\lambda_5+6g_2\lambda_5 \\
& -8g_2\lambda_4-8g_1\lambda_4-6g_3\lambda_6+10\lambda_4^2+\tfrac{9}{2}\lambda_5^2+\tfrac{9}{2}\lambda_6^2 ) \ ,  \\
\beta_{\lambda_5} =& -\varepsilon \lambda_5 +S_3 (-8g_1 \lambda_4 +8 g_2 \lambda_4 + 12 \lambda_4 \lambda_5) \ , \\
\beta_{\lambda_6} =& -\varepsilon \lambda_6 +S_3 (-8g_3 \lambda_4 +12 \lambda_4 \lambda_6) \ ,
\end{split}
\end{equation}
where as usual $\{g_i\}$ denote the couplings associated with the disorder-induced interactions, while $\{\lambda_i\}$ refer respectively to the marginal operators in the pure theory $\{ s',t_{1111},t_{1222}\}$. 

It's noteworthy that, in the absence of disorder and restricting to $t_{1111}$ only, our result can be mapped to the usual analysis of the rank-four perturbation in the study of systems with cubic perturbations. In a lattice model setup, this is the single-ion contribution \cite{pelissetto2002critical}. On the other side, in the field-theoretic description, this corresponds to the cubic-symmetric perturbation $\Phi \equiv \sum_i \phi_i^4$. This interaction is clearly not invariant under rotations, indicating that the model has the reduced symmetry of the lattice. In our notation, we are inserting
\begin{equation} 
t_{1111}=\frac{1}{2} \Phi -\frac{3}{2} s' \ ,
\end{equation}
thus the $O(2)$ symmetry is explicitly broken to a discrete cubic symmetry consisting of reflections and permutations of the field components. 

It is well known that the $N$-component Ginzburg-Landau model with a cubic anisotropy has four fixed points   (\cite{PhysRevB.8.4270} with \cite{Kleinert:1994td, carmona2000n} having pushed the accuracy of previous results to higher orders). These are the trivial Gaussian FP (free theory), the Ising FP (when the components of $\phi_i$ decouple), the $O(N)$-symmetric fixed point (also called Heisenberg FP), and the cubic FP. 
A special feature of the $O(2)$ case is the degeneracy in the critical exponents of the cubic and Ising fixed points. 

Our results agree with the literature and we can read off the fixed-points:
\begin{equation}
	\begin{split}
		& g_{s'}^G = 0 \ \ , \ \ g_{\Phi}^G=0 \ \ \ \ \ \ \ \ \ \ \ \ \ \ \ \  \rm{Gaussian \ FP} \ , \\
		& g_{s'}^H = \frac{\varepsilon}{20 \pi^2} \ \ , \ \ g_{\Phi}^H=0 \ \ \ \ \ \ \ \ \ \ \rm{Symmetric \ FP} \ , \\
		& g_{s'}^I = 0 \ \ , \ \ g_{\Phi}^I=\frac{\varepsilon}{72 \pi^2} \ \ \ \ \ \ \ \ \ \ \ \rm{Ising \ FP} \ , \\
		& g_{s'}^C = \frac{\varepsilon}{12 \pi^2} \ \ \ \ g_{\Phi}^C=-\frac{\varepsilon}{72 \pi^2} \  \ \ \  \rm{Cubic \ FP} \ .	
	\end{split}
\end{equation}
Let us now consider the insertion of $t_{1222}$ as well. This breaks the permutation symmetry but preserves $\mathbb{Z}_2$. From the beta-functions,
\begin{equation}
	\begin{split}
		& \beta_{s'}=-\varepsilon g_{s'} + 20 \pi^2 g_{s'}^2+48 \pi^2 g_{s'} g_{\Phi} + 9 \pi^2 \lambda_6^2 + \dots \ , \\
		& \beta_{\Phi}=-\varepsilon g_{\Phi} + 72 \pi^2 g_{\Phi}^2+24 \pi^2 g_{s'} g_{\Phi} + \dots \ , \\
		& \beta_{\lambda_6}=-\varepsilon \lambda_6 +72 \pi^2 g_{\Phi} \lambda_6 + 24 \pi^2 g_{s'}\lambda_6  + \dots \ ,
	\end{split}
\end{equation}
if $\lambda_6 \ne 0$, the one-loop approximation exhibits a degeneracy for $g_{\Phi}^*<\tfrac{\varepsilon}{72 \pi^2}$. In the simplest case of $g_\Phi^*=0$, we find:
\begin{equation}
	g_{s'}^* = \frac{\varepsilon}{24 \pi^2} \ , \ \lambda_6^* = \pm \frac{\varepsilon}{36 \pi^2} \ .
\end{equation}
If we finally include the disorder-induced interactions, we do not find any new fixed point and conclude that this charge-$2$ anisotropic random perturbation is irrelevant close to $4d$.

\section{Discussion and outlook}
\label{Conclusions}
In this work, we used the formalism of CPT to expand around a non-trivial fixed point and verify whether the introduction of a quenched disorder could alter its universality class. We addressed this study for the $O(2)$ critical model. 

First, we presented the random temperature perturbation, which is found to be relevant close to $4d$ but becomes irrelevant in $3d$. Then, we considered an anisotropic deformation associated with the insertion of $t_{ij}$ with a random coupling, potentially leading to the breaking of the $O(2)$ global symmetry. We found that this actually depends on which components of the disorder-induced interaction are considered. In three dimensions, if at the beginning we consider only the disorder preserving the global symmetry, the RG flow does not switch on the symmetry-breaking components of the disorder, resulting in a weakly coupled new fixed point in the infrared! Conversely, close to four dimensions, this quenched disorder is irrelevant.

Further steps could involve characterizing the theory at the random fixed point, for instance by studying the stress tensor operator to investigate the locality of the new theory. Additionally, since our approach involves analytic continuation in $n$ via the Replica Trick, it is important to verify whether reflection-positivity is preserved. Generically this method leads to a logarithmic CFT \cite{Hogervorst:2016itc}. 

Computing anomalous dimensions of operators within the same conformal multiplet or investigating the presence of a conserved current with no anomalous dimension could provide valuable insights as well. More generally, we would like to understand how to realize this disorder in a concrete model, perhaps by restricting to the case of the $O(2)$ preserving coupling.

Lastly, while our computations are valid at one-loop, extending them beyond the leading order to study observables in the vicinity of the fixed point requires knowledge of four-point correlators \cite{Bertucci:2022ptt}. For the $3d$ computation, the numerical bootstrap gives the conformal block expansion of the four-point functions. It is important to stress that the related two-loop beta-functions would still be scheme-independent, while the anomalous dimensions are generally scheme-dependent if they vanish at one-loop.

\section*{Acknowledgements }
I would like to express my deep gratitude to Alessandro Vichi for early collaborations on this project. A special thanks to Connor Behan and Luciano Viteritti, for stimulating discussions and comments on the draft. I am also grateful to Palash Singh, Adam Kmec, and Matteo Sacchi for helpful conversations.

\bibliography{references} 
\bibliographystyle{utphys}

\end{document}